\documentclass[letterpaper, 10 pt, conference]{ieeeconf}  % Comment this line out
\IEEEoverridecommandlockouts                              % This command is only
\newcommand{\CRIEEE}{©\the\year IEEE. Personal use of this material is permitted. Permission from IEEE must be obtained for all other uses, in any current or future media, including reprinting/republishing this material for advertising or promotional purposes, creating new collective works, for resale or redistribution to servers or lists, or reuse of any copyrighted component of this work in other works.}

\newcommand{\CRELSE}[1]{©\the\year. This manuscript version is made available under the CC-BY-NC-ND 4.0 license {#1}}

\newcommand{\CRIFAC}[2]{©\the\year { #1}. This work has been accepted to IFAC for publication under a Creative Commons Licence CC-BY-NC-ND {#2}}

\usepackage{graphics} % for pdf, bitmapped graphics files
\usepackage{epsfig} % for postscript graphics files
\usepackage{mathptmx} % assumes new font selection scheme installed
\usepackage{times} % assumes new font selection scheme installed
\usepackage{amsmath} % assumes amsmath package installed
\usepackage{amssymb}  % assumes amsmath package installed

\title{\LARGE \bf
Tracking of stabilizing, optimal control in fixed-time based on time-varying objective function
}

\author{Patrick Schmidt, Thomas Göhrt, and Stefan Streif% <-this % stops a space
\thanks{This work was partially funded by the European Union, European Social Found ESF, Saxony.}% <-this % stops a space
\thanks{The authors are with Technische Universität Chemnitz, Automatic Control and System Dynamics Lab, 09126 Chemnitz, Germany.
        {\tt\small \{patrick.schmidt, thomas.goehrt, stefan.streif\}@etit.tu-chemnitz.de}%
}
\thanks{\CRIEEE}
}

\usepackage{graphicx}
\usepackage{textcomp}

\usepackage{pgfplots}

\usepackage{float}
\usepackage{tikz}
\usetikzlibrary{arrows,%
                plotmarks, calc}
\usetikzlibrary[positioning, calc, intersections, backgrounds]

\usepackage[utf8]{inputenc}
  
\usepackage{xcolor}
\usepackage{xspace}
\usepackage{soul}
\usepackage{mathtools}
\usepackage{makecell}
\usepackage{cite}
\usepackage{nicefrac}
\usepackage{tcolorbox}
\usepackage{tcolorbox}

\newcommand{\ra}{\rightarrow}

\newcommand{\ie}{\unskip, i.\,e.,\xspace}
\newcommand{\eg}{\unskip, e.\,g.,\xspace}

\newcommand{\wrt}{w.\,r.\,t.\xspace}

\newcommand{\R}{\ensuremath{\mathbb R }}

\newcommand{\sm}{\ensuremath{\setminus}}

\newcommand{\red}[1]{\textcolor{red}{#1}}
\newcommand{\blue}[1]{\textcolor{blue}{#1}}

\definecolor{dgreen}{rgb}{0.0, 0.5, 0.0}

\DeclareMathOperator*{\argmin}{arg\,min}

\newcommand{\norm}[1]{\left\lVert#1\right\rVert}  % vector norm, needs ams
\newcommand{\abs}[1]{\left\lvert#1\right\rvert}

\makeatletter
\newcommand{\subalign}[1]{%
	\vcenter{%
		\Let@ \restore@math@cr \default@tag
		\baselineskip\fontdimen10 \scriptfont\tw@
		\advance\baselineskip\fontdimen12 \scriptfont\tw@
		\lineskip\thr@@\fontdimen8 \scriptfont\thr@@
		\lineskiplimit\lineskip
		\ialign{\hfil$\m@th\scriptstyle##$&$\m@th\scriptstyle{}##$\crcr
			#1\crcr
		}%
	}
}
\makeatother

\newtheorem{thm}{Theorem}
\newtheorem{dfn}{Definition}
\newtheorem{lem}{Lemma}

\newtheorem{asm}{Assumption}
\newtheorem{rem}{Remark}

\begin{document}

\maketitle
\thispagestyle{empty}
\pagestyle{empty}

%%%%%%%%%%%%%%%%%%%%%%%%%%%%%%%%%%%%%%%%%%%%%%%%%%%%%%%%%%%%%%%%%%%%%%%%%%%%%%%%
\begin{abstract}

The controller of an input-affine system is determined through minimizing a time-varying objective function, where stabilization is ensured via a Lyapunov function decay condition as constraint.
This constraint is incorporated into the objective function via a barrier function.
The time-varying minimum of the resulting relaxed cost function is determined by a tracking system.
This system is constructed using derivatives up to second order of the relaxed cost function and improves the existing approaches in time-varying optimization.
Under some mild assumptions, the tracking system yields a solution which is feasible for all times, and it converges to the optimal solution of the relaxed objective function in a user-defined fixed-time.
%the tracking error converges to zero in a user-defined fixed time and the solution of the tracking system is feasible for all times.
% and remains zero due to the tracking system.
The effectiveness of these results in comparison to exponential convergence is demonstrated in a case study.

%\begin{itemize}
%	\item \red{description of the problem addressed (1-2 sentences)}
%	\item \red{statement of the main result (1-2 sentences)}
%	\item \red{key conclusion that you want to highlight (1-2 sentences)}
%	\item \red{comment on an application/numerical example included (1-2 sentences)}
%\end{itemize}
\end{abstract}

%%%%%%%%%%%%%%%%%%%%%%%%%%%%%%%%%%%%%%%%%%%%%%%%%%%%%%%%%%%%%%%%%%%%%%%%%%%%%%%%
\section{INTRODUCTION} \label{sec:intro}

Consider the following input-affine system
\begin{equation} \label{eqn:dynamics-f}
	\dot x = f(x) + g(x) u
\end{equation}
with states $x \in \R^n$, controls $u \in \R^m$ and initial condition $x(0) = x_0$.
The internal dynamics model $f: \R^n \ra \R^n$ and the input coupling function $g: \R^n \ra \R^{n \times m}$ are assumed to be continuously differentiable in $x$.

A stabilizing control for \eqref{eqn:dynamics-f} can be determined by minimizing a user-defined objective function, which varies in time.
Time-varying objective functions appear in many applications due to a dynamic environment\eg in autonomous driving or unmanned aerial vehicles, where changing weather conditions affect possible trajectories and system states such as battery capacity. 
Therefore, the solution of the resulting optimization problem is also time-dependent.

There exist several ways to identify an optimal control \eg model predictive control (MPC) %, where a cost function is optimized over a moving horizon 
\cite{Allgower2009-nonlinear}, \cite{Fontes2001-general}, \cite{Gruene2017-nonlinear}, \cite{Zavala2009-advanced}, adaptive dynamic programming (ADP) %, which  infinite-horizon optimal control problem
\cite{Goehrt2019-adaptive-actor-critic}, \cite{Goehrt2019-adaptive}, \cite{Jiang2015-global}, \cite{Liu2017-adaptive}, or time-varying optimization (TVO) \cite{Bernstein2019-online}, \cite{Fazlyab2017-prediction}, \cite{Simonetto2018-dual}, \cite{Simonetto2015-non-diff-TVO}.
The latter method is considered in the paper, since it is based on the continuous-time system \eqref{eqn:dynamics-f} instead of a discretized model, and thus it yields a continuous-time optimal control.
In TVO, a dynamical system (the so-called tracking system) is constructed, which tracks the optimal solution of the given objective function.
The convergence to the optimal solution can be adapted based on the design of the tracking system.
Since a non-optimal control causes higher costs, it is important, especially from an economic point of view, to identify the optimal solution preferably fast, but taking limitations of control actions into account.
Therefore, the goal of this paper is the design of a tracking system, whose solution converges to the optimal solution in a user-defined time.
Furthermore, this convergence time should be independent of the initial value of the tracking system, to guarantee a convergence at the specified time.
These results can be applied on many application areas, such as communications, control systems, cyber-physical systems, machine learning, medical engineering, process engineering, and robotics.
An overview of TVO in these application areas can be found in \cite{Simonetto2020-time-varying-survey}.

%\begin{itemize}
%	\item \red{First paragraph: briefly describe the problem (expanded version of the abstract)}
%	\item \red{Second (and maybe further) paragraph: relevant literature}
%	\item \red{Next paragraphs: contribution of the paper (skip unimportant details; focus readers attention to the most important achievements; link to literature)}
%	\item \red{Last paragraph: Remainder of the paper}
%\end{itemize}

%\red{emph nutzen, wenn Begriffe zum ersten Mal eingeführt werden. Überlegen, ob man jedes Remark braucht}

In discrete-time TVO, the optimization problem is sampled at different time instants, which results in a sequence of time-invariant optimization problems.
The solution at the next time step is predicted based on the current one and then corrected.
The solution of the prediction-correction system converges $Q$-linearly (the discrete-time counterpart of exponential convergence) to the sequence of optimal controls \cite{Simonetto2020-time-varying-survey}.
In \cite{Bastianello2019-prediction}, \cite{Simonetto2017-first-order-pred-corr}, \cite{Simonetto2017-pred-corr-constr}, \cite{Simonetto2016-pred-corr} different settings by means of state constraints, non-differentiable cost functions, and prediction-correction algorithms are considered.
In continuous-time, TVO is considered in \cite{Fazlyab2016-self-triggered}, \cite{Fazlyab2016-interior}, \cite{Fazlyab2017-prediction}, \cite{Fazlyab2018-analysis}, as well as in \cite{Garg2020-prescribed-time}, \cite{Song2019-time-varying-feedback}, \cite{Udwadia2008-optimal}, \cite{Zheng2020-implicit}.
These works capture the idea of discrete-time TVO and show exponential convergence of the tracking error, which is sufficient in many applications.
Nevertheless, the tracking error in \cite{Fazlyab2017-prediction} does not converge in finite time and it depends on the initial value of the tracking system.
This paper extends these results to fixed-time convergence, which yields a vanishing tracking error after a user-defined time.
This time is independent of the initial value of the tracking system.
Fixed-time convergence was considered in many settings, such as observer design \cite{Menard2017-fixed-time-observer}, parameter estimation \cite{Rios2017-time-varying-parameter}, sliding mode control \cite{Levant2013-fixed-time-SMC}, extremum seeking \cite{Poveda2020-fixed}, and also controller design \cite{Polyakov2011-nonlinear-feedback}, \cite{Polyakov2019-consistent-discr}, \cite{Polyakov2015-implicit-LF}.
The latter publications provide differential equations, whose solutions converge in fixed-time to zero.
In this work, these results on fixed-time convergence are generalized and applied to TVO to construct a tracking system, which yields a feasible solution that converges to the minimizer of the relaxed objective function and ensures a vanishing tracking error, which distinguishes it from other existing approaches.

The paper is structured as follows: The preliminaries and the problem setting are defined in Section \ref{sec:problem-setting}.
Section \ref{sec:main-results} contains the main result, namely the design of a tracking system whose solution converges to the optimal solution in fixed-time and stabilizes \eqref{eqn:dynamics-f}.
A case study is presented in Section \ref{sec:case-study}, which compares the proposed approach with an existing one that ensures exponential convergence.
Finally, Section \ref{sec:conclusion} concludes the paper and provides an outlook on further research topics.

Notation: Let $\R_{>0}$ be defined as $\R_{>0} := \{ z \in \R: z  > 0 \}$; $\R_{\geq 0}$, $\R_{\leq 0}$ and $\R_{< 0}$ are defined analogously.
Furthermore, $\norm{x}$ denotes the Euclidean norm of a vector $x$.%, and $\norm{X}$ denotes the spectral norm of a matrix $X$. 
%\blue{Lastly, $\nabla_x J$ describes the Jacobian of a function $J$ w.r.t. $x$, whereas $\nabla_{xx} J$ is the Hessian matrix of $J$. }

%%%%%%%%%%%%%%%%%%%%%%%%%%%%%%%%%%%%%%%%%%%%%%%%%%%%%%%%%%%%%%%%%%%%%%%%%%%%%%%%

\section{PROBLEM SETTING} \label{sec:problem-setting}

Consider the following optimization problem:
\begin{equation}
	\label{eqn:opt-prob-initial}
	\begin{split}
		\min_{u \in \R^m} \quad&J(u,x)  \\
		\mathrm{s.t.}\quad& \dot x = f(x) + g(x)u, \\
		&\nabla_x V(x)^\top [f(x) + g(x) u] \leq - w(x),
	\end{split}
\end{equation}
where $\nabla_x V(x)$ is denoted as the Jacobian of $V$.
The objective function $J: \R^m \times \R^n \ra \R$ is assumed to satisfy the following standard assumptions (cf. \cite{Fazlyab2016-self-triggered}, \cite{Simonetto2016-pred-corr}):

\begin{asm} \label{asm:J-twice-differentiable}
	The objective function $J(u,x)$ in \eqref{eqn:opt-prob-initial} is twice continuously differentiable in both arguments.
\end{asm}

\begin{asm} \label{asm:J-uni-str-convex}
	The objective function $J(u,x)$ in \eqref{eqn:opt-prob-initial} is uniformly strongly convex in $u$ with parameter $m_J$ for all $x \in \R^n$ \ie for the Hessian of $J$, which is denoted as $\nabla_{uu} J$ holds: $\nabla_{uu} J(u,x) - m_J I$ is positive semidefinite for all $u \in \R^m$ and all $x \in \R^n$.
\end{asm}

The existence of a stabilizing control $u$ for \eqref{eqn:dynamics-f} is ensured through the following assumption:

\begin{asm} \label{asm:decay-condition}
	There exists a continuously differentiable feedback controller $\kappa(x)$ and a twice continuously differentiable radially unbounded Lyapunov function $V$ such that for all $x \in \R^n$, it holds that 
	\begin{equation} \label{eqn:decay-condition}
	\dot V(x) = \nabla_x V(x)^\top \left[ f(x) + g(x) \kappa(x) \right] \leq - w(x), 
	\end{equation}
	where $w: \R^n \ra \R$ is continuously differentiable and satisfies $w(0) = 0$ and $x \not = 0 \implies w(x) > 0$.
\end{asm}

Assumption \ref{asm:J-twice-differentiable} is used for the main theorem in Section \ref{sec:main-results} and Assumption \ref{asm:J-uni-str-convex} implies that the solution trajectory of \eqref{eqn:opt-prob-initial} is unique for all $t \in \R_{\geq 0}$, whereas their existence is ensured via Assumption \ref{asm:decay-condition}.

Let the inequality constraint in \eqref{eqn:opt-prob-initial} be defined as
\begin{equation} \label{eqn:varphi}
	\varphi(u,x) := \nabla_x V(x)^\top [f(x) + g(x) u] + w(x). 
\end{equation}
Instead of satisfying $\varphi(u,x) \leq 0$ for all $x \in \R^n, u \in \R^m$, the inequality is relaxed to $\varphi(u,x) \leq \gamma$ for $\gamma > 0$.
This relaxed inequality is included in the objective function of optimization problem \eqref{eqn:opt-prob-initial} via a barrier function (cf. \cite{Fazlyab2017-prediction}, \cite{Goehrt2019-adaptive}):

\begin{dfn}[Barrier function] \label{dfn:barrier-fct}
	A twice continuously differentiable convex function $B: \R_{\leq 0} \ra \R \cup \{ \infty \}$ is called barrier function, if $B(z) < \infty$ for all $z < 0$ and $\lim\limits_{z \ra 0^-} B(z) \ra \infty$.
\end{dfn}

The resulting relaxed optimization problem is given as
\begin{equation}
	\label{eqn:opt-prob}
	\begin{split}
		\min_{u \in \R^m} \quad& \tilde J(u,x,t) \\
		\mathrm{s.t.}\quad& \dot x = f(x) + g(x) u,
	\end{split}
\end{equation}
where the relaxed objective function $\tilde J(u,x,t)$ is defined as
\begin{equation} \label{eqn:dfn-of-J-tilde}
	\tilde J(u,x,t) := J(u,x) + \mu(t) B \left( \varphi(u,x) - \gamma \right)
\end{equation}
and $\mu(t)$ is continuously differentiable and satisfies $\mu(t) > 0$ and $\dot \mu(t) < 0$ for all $t \in \R_{\geq 0}$.
The relaxed inequality ensures a well-defined optimization problem even if $\varphi(u,x) = 0$, which occurs at $x = 0$.

Due to Lemma \ref{lem:J-tilde-strongly-convex} in the appendix, $\tilde J(u,x,t)$ is strongly convex with parameter $m_J$, since it is the sum of the strongly convex function $J(u,x)$ and the convex function $\mu(t) B(\varphi(u,x) - \gamma)$.
This means that the solution of \eqref{eqn:opt-prob} is unique for all $t \in \R_{\geq 0}$.
The cost function $J$ often involves a quadratic weighting in the input, like $J = u^\top R u$, where a positive definite matrix $R$ yields the required strong convexity \cite{Bastianello2019-prediction}, \cite{Fazlyab2016-interior}, \cite{Romero2020-continuous-time}.

The optimal input is defined as
\begin{equation}
	\label{eqn:opt-prob-argmin}
	\begin{split}
		u^\star(t) = \argmin_{u \in \R^m} \quad&\tilde J(u,x,t) \\
		\mathrm{s.t.}\quad& \dot x = f(x) + g(x) u.
	\end{split}
\end{equation}
As described in the introduction, methods from time-varying optimization are applied to track $u^\star(t)$.
Therefore, a dynamical system 
\begin{equation}
	\label{eqn:system-u}
	\dot u = h(u, x, t), \ u(0) = \kappa(x_0),
\end{equation}
is constructed in Section \ref{sec:main-results}, whose solution $u(t)$ converges to $u^\star(t)$.

%%%%%%%%%%%%%%%%%%%%%%%%%%%%%%%%%%%%%%%%%%%%%%%%%%%%%%%%%%%%%%%%%%%%%%%%%%%%%%%%

\section{MAIN RESULTS} \label{sec:main-results}

The design of \eqref{eqn:system-u} is a crucial element for a vanishing tracking error $e(t) := \norm{u(t) - u^\star(t)}^2$.
To make this error quickly converge to zero independent of the initial value in \eqref{eqn:system-u}, a tracking system that guarantees fixed-time convergence is constructed.

\subsection{Fixed-time convergence} \label{subsec:fixed-time-conv}

Fixed-time stability is defined as follows \cite{Polyakov2011-nonlinear-feedback}, \cite{Polyakov2015-implicit-LF}:

\begin{dfn}[Fixed-time stability] \label{dfn:fixed-time}
	The origin of \eqref{eqn:system-u} is said to be globally fixed-time stable, if it is globally uniformly asymptotically stable \cite{Khalil2002-nonlinear}	and there exists a settling-time $T \geq 0$, such that for all $u_0 \in \R^m$ it holds that $\norm{u(t)} = 0$ for all $t \geq T$, where $u(\cdot)$ is an arbitrary solution of \eqref{eqn:system-u}.
\end{dfn}

\begin{rem}[Finite-time stability]
	On the contrary to Definition \ref{dfn:fixed-time}, the settling-time in finite-time stability depends on the initial value $u(0)$.
\end{rem}

%Note that fixed-time stability is also known as prescribed-time stability \cite{Garg2019-prescribed-time}. 
Examples for finite-time and fixed-time stable systems are given in \cite{Polyakov2011-nonlinear-feedback}, \cite{Polyakov2015-implicit-LF}.
For the following analysis, 
\begin{equation} \label{eqn:ODE-fxt-conv}
	\dot z = - \frac \pi \tau (\abs z ^{0.5} + \abs z^{1.5}) \text{sign}(z), \ z(0) = z_0
\end{equation}
with $z \in \R$ is considered.

\begin{lem}[Fixed-time stable system] \label{lem:fixed-time-system}
	Consider \eqref{eqn:ODE-fxt-conv} and let $\tau \in \R_{> 0}$ be the user-defined settling-time.
	The origin of \eqref{eqn:ODE-fxt-conv} is fixed-time stable with settling-time $T = \tau$.
	Furthermore, the solution is given as
	\begin{equation}\label{eqn:ODE-fxt-conv-solution}
		z(t) = \text{sign}(z_0) \left[ \tan\left(\arctan \left( \sqrt{\abs{z_0}} \right) - \frac{\pi t}{2 \tau} \right) \right]^2.
		% =: \varphi(t, z_0).
	\end{equation}
 \end{lem}

\begin{proof}
	The solution of \eqref{eqn:ODE-fxt-conv} is achieved via separation of variables. 
	Since $z(t) = 0$ for all $t \geq T$, the settling-time $T$ is computed using the solution \eqref{eqn:ODE-fxt-conv-solution} as:
	\begin{equation} \label{eqn:settling-time-T-pi}
		T = \sup_{z_0 \in \R} \left\{ 2 \frac{\tau}{\pi} \arctan(\sqrt{\abs{z_0}}) \right\} = \tau.
	\end{equation}
\end{proof}

Since the solution of \eqref{eqn:ODE-fxt-conv} might not be unique due to the discontinuous right-hand side of \eqref{eqn:ODE-fxt-conv}, it has to be defined as Filippov solution \cite{Filippov1988-differential}.
Please refer to \cite{Polyakov2015-implicit-LF}, \cite{Romero2020-continuous-time} or \cite{Romero2020-finite-time} for further details.

\begin{rem}[Choice of settling-time $\tau$] \label{rem:user-defined-settling-time}
	Lemma \ref{lem:fixed-time-system} demonstrates that the convergence of \eqref{eqn:ODE-fxt-conv-solution} to zero can be made arbitrarily fast, which results in a high gain.
	This may cause numerical problems (cf. \cite[Remark 1]{Fazlyab2017-prediction}), which is also described in the case study in Section \ref{sec:case-study}.
\end{rem}

In the following subsection, differential equation \eqref{eqn:ODE-fxt-conv} is used to construct a tracking system \eqref{eqn:system-u}.

%%%%%%%%%%%%%%%%%%%%%%%%%%%%%%%%%%%%%%%%%%%%%%%%%%%%%%%%%%%%%%%%%%%%%%%%%%%%%%%%
\subsection{Construction of tracking system} \label{subsec:main}

To identify $u^\star(t)$ in \eqref{eqn:opt-prob-argmin}, the following tracking system is proposed:
\begin{equation} \label{eqn:special-choice-h}
	\begin{aligned}	
		\dot u(t) = &- \nabla_{uu}^{-1} \tilde J(u, x, t) \left[ \Psi(\nabla_u \tilde J(u, x, t)) + \nabla_{ut} \tilde J(u, x, t) \right. \\
		&+ \left.\nabla_{ux} \tilde J(u, x, t)^\top \left[ f(x) + g(x) u \right] \right], u(0) = \kappa(x_0),
	\end{aligned}
\end{equation}
where $\Psi: \R^m \ra \R^m$ is defined as
\begin{equation} \label{eqn:G-fixed-time-functions-vector}
	\begin{aligned}
		&\Psi(u) := \begin{pmatrix}
			\psi(u_1) \\ \psi(u_2) \\ \vdots \\ \psi(u_m)
		\end{pmatrix}, \psi(u_i) = \frac \pi \tau (\abs{u_i}^{0.5} + \abs{u_i}^{1.5}) \text{sign}(u_i).
	\end{aligned}
\end{equation}
In the following Theorem, it is shown that $u(t)$ as the solution of \eqref{eqn:special-choice-h} is feasible for all $t \in \R_{\geq 0}$, asymptotically stabilizes \eqref{eqn:dynamics-f}, and converges to $u^\star(t)$ defined in \eqref{eqn:opt-prob-argmin} in fixed-time with settling-time $T =  \tau$.
From now on, function arguments are omitted for a clearer presentation of the results.

%
%\begin{lem}[Feasibility] \label{lem:feasibility}
%	Consider system \eqref{eqn:dynamics-f}, optimization problem \eqref{eqn:opt-prob}, update rule \eqref{eqn:special-choice-h} and let Assumptions \ref{asm:J-twice-differentiable}-\ref{asm:decay-condition} hold.
%	Define $\Phi(x) := \{ u \in \R^m: \varphi(u,x) \leq \gamma \}$.
%	Then, for every initial condition $x_0 \in \R^n$, $u(t)$ as a solution of \eqref{eqn:special-choice-h} satisfies $u(t) \in \Phi(x(t))$ for all $t \in \R_{\geq 0}$, if $u(0) \in \Phi(x(0))$.
%\end{lem}
%
%\begin{proof}
%	 
%\end{proof}

\begin{thm}[Fixed-time convergence] \label{thm:fixed-time-convergence-TVO}
	Consider the system \eqref{eqn:dynamics-f}, optimization problem \eqref{eqn:opt-prob}, update rule \eqref{eqn:special-choice-h} and let Assumptions \ref{asm:J-twice-differentiable}, \ref{asm:J-uni-str-convex}, and \ref{asm:decay-condition} hold.
	Let $u^\star(t)$ be given as \eqref{eqn:opt-prob-argmin} and $u(t)$ be given as the solution of \eqref{eqn:special-choice-h}.
	Furthermore, define the set of feasible controls as $\Phi(x) := \{ u \in \R^m: \varphi(u,x) \leq \gamma \}$.	
	Then, $u(t)$ as a solution of \eqref{eqn:special-choice-h} with initial value $u(0) \in \Phi(x(0))$ satisfies $u(t) \in \Phi(x(t))$ for all $t \in \R_{\geq 0}$ and for every initial condition $x_0 \in \R^n$.
	Furthermore, the tracking error $\norm{u(t) - u^\star(t)}$ converges to zero in fixed-time with settling-time $T = \tau$.
\end{thm}

\begin{proof}
The proof is divided into two parts: In the first part, feasibility of $u(t)$ is shown.
The second part shows the fixed-time convergence of $\norm{u(t)-u^\star(t)}$.

	\underline{Part 1:} Feasibility: \newline
	It is shown that $u(t)$ as a solution of \eqref{eqn:special-choice-h} is feasible for all $t \geq 0$ \ie $\varphi(u,x) \leq \gamma$ for all $t \in \R_{\geq 0}$.
	Therefore, the time derivative of $\nabla_u \tilde J$ at $(u(t),x(t),t)$ is considered and $\dot u(t)$ is replaced by \eqref{eqn:special-choice-h}:
	\begin{equation} \label{eqn:system-nabla-J-unconstr}
	\begin{aligned}
		&\dot \nabla_u \tilde J(u,x,t) \\
		&= \nabla_{uu} \tilde J(u,x,t) \dot u(t) + \nabla_{ux} \tilde J(u,x,t)^\top \dot x(t) + \nabla_{ut} \tilde J(u,x,t) \\
		&= - \Psi(\nabla_u \tilde J(u,x,t)), \ \nabla_u \tilde J(u(0),x(0),0) = \nabla_u \tilde J(\kappa(x_0),x_0,0).
	\end{aligned}
	\end{equation}
	The solution of \eqref{eqn:system-nabla-J-unconstr} is given as
	\begin{equation} \label{eqn:solution-nabla-J}
		\nabla_u \tilde J(u,x,t) = U\left( t \right),
	\end{equation}
	where $U$ is defined as
	\begin{equation}
		\begin{split}
			U \left( t \right) = \begin{pmatrix}
				\text{sign} \left( \zeta_{0,1} \right) \left[ \tan \left( \arctan \left( \sqrt{\abs{\zeta_{0,1}}} \right) - \frac{\pi t}{2 \tau} \right) \right]^2 \\
				\vdots \\
				\text{sign} \left( \zeta_{0,m} \right) \left[ \tan \left( \arctan \left( \sqrt{\abs{\zeta_{0,m}}} \right) - \frac{\pi t}{2 \tau} \right) \right]^2
			\end{pmatrix}
		\end{split}
	\end{equation}
	and  $\zeta_0 = \begin{pmatrix} \zeta_{0,1} & \cdots & \zeta_{0,m} \end{pmatrix} := \nabla_u \tilde J(u(0),x(0),0) \in \R^m$ is the initial value.
	
	The mean-value theorem is used to expand the solution \eqref{eqn:solution-nabla-J} \wrt $u$ around $u^\star(t)$ as
	\begin{equation} \label{eqn:mean-value-thm}
		\begin{aligned}
			&\nabla_u \tilde J(u,x,t) = \nabla_{uu}\tilde J(v,x,t) (u(t) - u^\star(t)) \\
			&\Leftrightarrow u(t) - u^\star(t) = \nabla_{uu}^{-1}\tilde J(v,x,t) \nabla_u \tilde J(u,x,t).
		\end{aligned}
	\end{equation}
	Note that $v = v(t)$ is a convex combination of $u(t)$ and $u^\star(t)$.
	Furthermore, $\nabla_u \tilde J(u^\star,x,t) = 0$ holds due to the first-order optimality condition.
	
	Consider \eqref{eqn:mean-value-thm} and equation \eqref{eqn:solution-nabla-J}.
	It follows that
	\begin{equation} \label{eqn:convergence-x-x-star-proof}
		\begin{split}
			&\norm{u(t) - u^\star(t)} \\
			&\leq \norm{\nabla_{uu}^{-1}\tilde J(v,x,t)} \norm{\nabla_u \tilde J(u,x,t)} \leq \frac{1}{m_J} \norm{ U(t) }.
		\end{split}
	\end{equation}
	Furthermore, with \eqref{eqn:convergence-x-x-star-proof} and the strong convexity of $\tilde J(u,x,t)$ it follows that
	\begin{equation} \label{eqn:convergence-Jx-Jx-star-proof}
		\begin{split}
			&\underbrace{\tilde J(u,x,t) - \tilde J(u^\star,x,t)}_{\geq 0} \leq \nabla_u \tilde J(u,x,t)^\top (u(t) - u^\star(t)) \\
			&\leq \norm{ \nabla_u \tilde J(u,x,t) } \norm{u(t) - u^\star(t)} \leq \frac{1}{m_J} \norm{ U(t) }^2.
		\end{split}
	\end{equation}	
	Note that $u(0)$ is feasible due to the initial value in \eqref{eqn:special-choice-h}.
	Furthermore, $u(t)$ is feasible for $t \geq \tau$, since $u^\star(t)$ as a solution of \eqref{eqn:opt-prob} is feasible for all $t \in \R_{\geq 0}$ and the tracking error $\norm{u(t) - u^\star(t)}$ is zero for $t \geq \tau$.
	Therefore, feasibility of $u(t)$ for $t \in (0,\tau)$ has to be shown, which is done indirectly.
	Assume that there exists a 
	\begin{equation}
		t_2 := \min_{t \in \R_{\geq 0}} \{ t: \varphi(u(t),x(t)) - \gamma = 0 \},
	\end{equation}
	which is the earliest time where $u(t)$ is not feasible.
	Since $t_2$ exists due to our assumption, there exists also a
	\begin{equation}
		t_1 := \max_{t \in [0,t_2)} \{ t: \varphi(u(t),x(t)) - \gamma < 0 \}.
	\end{equation}
	On the one hand, it holds that $\tilde J(u(t),x(t),t) < \infty$ for all $t \leq t_1$, since $u(t)$ is feasible for all $t \leq t_1$ based on the definition of $t_1$.
	On the other hand, since $\mu(t_2) > 0$ and $\lim_{t \ra t_2} B(\varphi(u(t),x(t)) - \gamma) = \infty$, the value of the objective function at $t_2$ is $\lim_{t \ra t_2} \tilde J(u(t),x(t),t) = \infty$.
	This is a contradiction to \eqref{eqn:convergence-Jx-Jx-star-proof}, since $\norm{U(t)}^2$ is monotony decreasing (cf. \eqref{eqn:ODE-fxt-conv-solution}). 
	Therefore, $u(t)$ is feasible for all $t \in \R_{\geq 0}$ and thus it stabilizes \eqref{eqn:dynamics-f} based on Assumption \ref{asm:decay-condition}.
	
	\underline{Part 2:} Fixed-time convergence: \newline
	Consider a candidate Lyapunov function 
	\begin{equation} \label{eqn:CLF-u}
		\tilde V(\nabla_u \tilde J) := \frac 1 2 \nabla_u \tilde J^\top \nabla_u \tilde J
	\end{equation}
	of \eqref{eqn:special-choice-h}.
	It is shown that \eqref{eqn:CLF-u} is a Lyapunov function (LF) for \eqref{eqn:special-choice-h}.
	Indeed $\tilde V(\nabla_u \tilde J)$ is positive definite for all $\nabla_u \tilde J \in \R^m$ and zero iff $\nabla_u \tilde J = 0$.
	Now, consider the derivative of $\tilde V$:
	\begin{equation} \label{eqn:V-dot-1}
		\begin{split}
			\dot{\tilde V}(\nabla_u \tilde J) &= \nabla_u \tilde J^\top \dot \nabla_u \tilde J \\
			&= \nabla_u \tilde J^\top \left[ \nabla_{uu} \tilde J \dot u + \nabla_{ux} \tilde J^\top \dot x + \nabla_{ut} \tilde J			\right] \\
			&= - \nabla_u \tilde J^\top \Psi(\nabla_u \tilde J),
		\end{split}
	\end{equation}
	where $\dot u(t)$ was replaced by \eqref{eqn:special-choice-h}.
	The last line in \eqref{eqn:V-dot-1} reads as
	\begin{equation} \label{eqn:V-dot-2}
		\begin{split}
			\dot{\tilde V} &= - \nabla_u \tilde J^\top \Psi(\nabla_u \tilde J) \\
			&= -\sum_{i = 1}^m \frac{\partial \tilde J}{\partial u_i} \psi\left( \frac{\partial \tilde J}{\partial u_i} \right) \\
			&= -\sum_{i = 1}^m \frac{\partial \tilde J}{\partial u_i} \left( \abs{\frac{\partial \tilde J}{\partial u_i}}^{0.5} + \abs{\frac{\partial \tilde J}{\partial u_i}}^{1.5} \right) \text{sign} \left( \frac{\partial \tilde J}{\partial u_i} \right) \\
			&=  -\sum_{i = 1}^m \abs{\frac{\partial \tilde J}{\partial u_i}} \left( \abs{\frac{\partial \tilde J}{\partial u_i}}^{0.5} + \abs{\frac{\partial \tilde J}{\partial u_i}}^{1.5} \right).
		\end{split}
	\end{equation}	
	Consider the last sum in \eqref{eqn:V-dot-2}. 
	Each element is greater than zero, and thus $\tilde V$ is the negative sum of these elements.
	Hence, $\dot{\tilde V} \leq 0$ for all $\nabla_u \tilde J \in \R^m$.
	Furthermore, $\dot{\tilde V} = 0$ holds iff each element $\frac{\partial \tilde J}{\partial u_i} = 0$ for all $i = 1, \ldots, m$.
	Therefore, $\dot{\tilde V}$ is zero iff $\nabla_u \tilde J = 0$, and thus $\tilde V$ is a LF for \eqref{eqn:special-choice-h}.
	Now, consider the second line in \eqref{eqn:V-dot-2}.
	Since $\psi\left( \frac{\partial \tilde J}{\partial u_i} \right) = 0$ for $t \geq \tau$ (due to its construction in Lemma \ref{lem:fixed-time-system}) and $\frac{\partial \tilde J}{\partial u_i} < \infty$ (since $u(t)$ is feasible), it follows that $\dot{\tilde V} = 0$ for all $t \geq \tau$ and for all $x \in \R^n$.
	Hence, $\dot{\tilde V} = 0$ implies $\nabla_u \tilde J = 0$, which holds iff $u(t) = u^\star(t)$, since $u^\star(t)$ is unique.
	Thus, $\norm{u(t) - u^\star(t)} = 0$ for all $t \geq \tau$ which proves fixed-time convergence of the tracking error to zero.
\end{proof}
The next section provides a case study for the suggested approach.

%%%%%%%%%%%%%%%%%%%%%%%%%%%%%%%%%%%%%%%%%%%%%%%%%%%%%%%%%%%%%%%%%%%%%%%%%%%%%%%%

\section{CASE STUDY AND DISCUSSION} \label{sec:case-study}

The performance of tracking system \eqref{eqn:special-choice-h} is investigated for the following system dynamics
\begin{equation} \label{eqn:system-case-study}
	\dot x = \begin{pmatrix}
		- x_1 - x_2^2 \\ 
		x_1 x_2 + x_2 x_3 \\
		-x_2^2 - x_3
	\end{pmatrix} + \begin{pmatrix}
		0 & 0 \\ 1 & 0 \\ 0 & 1
	\end{pmatrix} u.
\end{equation}
System \eqref{eqn:system-case-study} is an extension of the system which is considered in \cite{Goehrt2019-adaptive}.
It is stabilized by an optimal control given as the minimum of objective function
\begin{equation} \label{eqn:system-J}
	J(u) = \frac 1 2 (u_1^2 + 3 u_2^2).
\end{equation}
Let $V(x) = \nicefrac 1 2 \norm x^2$ be a control Lyapunov function (CLF) for \eqref{eqn:system-case-study}.
Its derivative is given as
\begin{equation}
	\begin{split}
		&\dot V(x) \\
		&= \nabla_x V(x)^\top (f(x)+g(x)u) \\
		&= x_1 (- x_1 - x_2^2) + x_2(x_1 x_2 + x_2 x_3 + u_1) + x_3(-x_2^2 - x_3 + u_2) \\
		&= -x_1^2 - x_3^3 + u_1 x_2 + u_2 x_3.
	\end{split}
\end{equation}
Therefore, both solutions $\kappa_1(x) = \begin{pmatrix} -(x_1+x_2), & -x_2 \end{pmatrix}^\top$ and $\kappa_2(x) = \begin{pmatrix} -(x_1+x_2+x_3), & 0 \end{pmatrix}^\top$ asymptotically stabilize \eqref{eqn:system-case-study}, since they yield
\begin{equation}
	\dot V(x) = - \norm x^2 - x_1x_2 - x_2x_3 = - x^\top \underbrace{\begin{pmatrix}
		1 & 0.5 & 0 \\ 0.5 & 1 & 0.5 \\ 0 & 0.5 & 1
	\end{pmatrix}}_{=: P} x,
\end{equation}
where $P$ is a positive definite matrix.
Thus, $\dot V(x) < 0$ for all $x \in \R^3 \sm \{ 0 \}$.
Based on $\dot V(x)$, the decay rate in \eqref{eqn:varphi} is defined as $w(x) = 0.1 \norm x^2$.
Furthermore, let $\gamma = 0.01$, $\mu(t) = \exp(-t)$, and $B(z) = - \nicefrac 1 z$ be given.
The settling-time is set to $\tau = 3 \text s$.
The initial values are considered as $x(0) = x_0 = \begin{pmatrix} -9, & -7, & -5 \end{pmatrix}^\top$ and $u(0) = u_0 = \kappa_1(x_0)$.

The performance of tracking system \eqref{eqn:special-choice-h} is compared with a tracking system proposed in \cite{Fazlyab2017-prediction}, which ensures exponential convergence (EC) of the tracking error.
In the case of EC, $\Psi(\nabla_u \tilde J(u, x, t))$ in \eqref{eqn:special-choice-h} is replaced by $A \nabla_u \tilde J(u, x, t)$, where $A$ is a positive definite matrix satisfying $\alpha I \preceq A \in \R^{m \times m}$.
For the following considerations, $A$ was set to $A = \text{diag}(1,1)$ in the computations.
To distinguish the solutions of the FC and EC approach, the derived control of the EC approach is defined as $\hat u(t)$, whereas the resulting states are defined as $\hat x(t)$, and the optimal control is given as $\hat u^\star(t)$.

\begin{figure}[thpb]
	\centering
    \includegraphics[width = 0.42\textwidth]{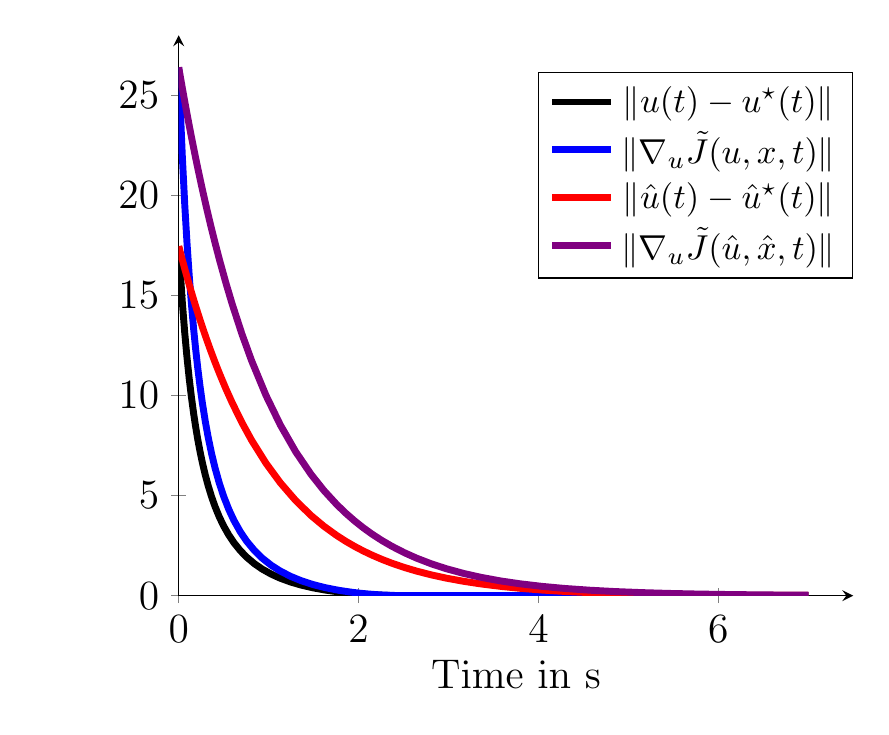}
    \caption{Exponential and fixed-time convergence of tracking error and gradient of objective function.}
    \label{fig:convergences}
\end{figure}

\begin{figure}[thpb]
	\centering
    \includegraphics[width = 0.42\textwidth]{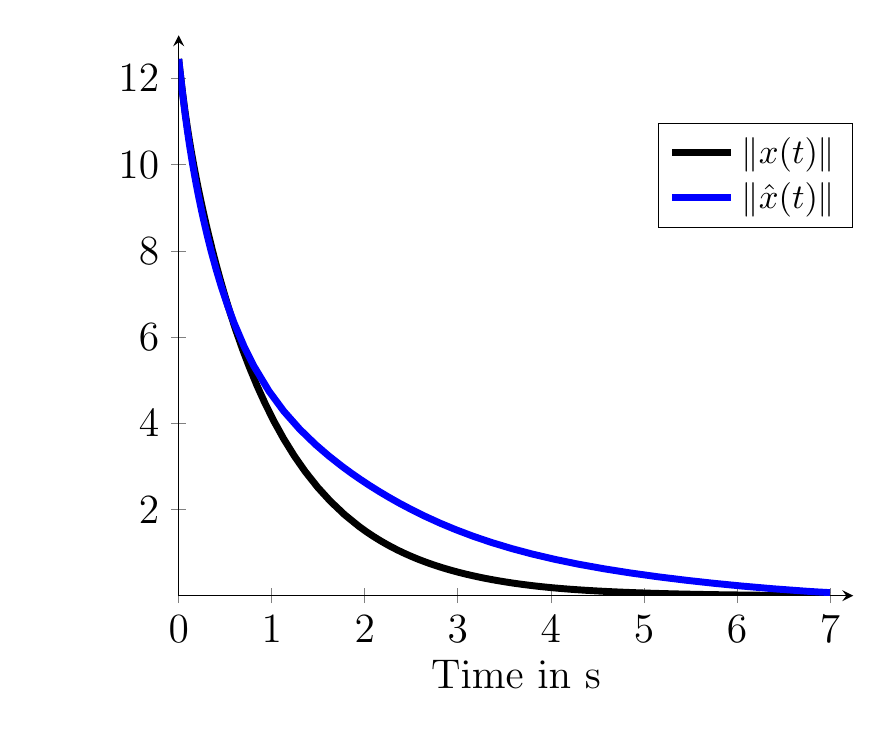}
    \caption{Convergence of states for exponential and fixed-time convergence.}
    \label{fig:convergence-states}
\end{figure}

Fig. \ref{fig:convergences} compares the convergence of the tracking error $\norm{u(t) - u^\star(t)}$ in the fixed-time convergence (FC) approach, or, respectively, $\norm{\hat u(t) - \hat u^\star(t)}$ in the EC approach.
Both tracking errors converge to zero, whereas the convergence of $\norm{u(t) - u^\star(t)}$ is faster and happens in fixed-time.
Also the gradient $\norm{\nabla_u \tilde J(u,x,t)}$ converges in this time, since it is zero iff $u(t) = u^\star(t)$.
The faster convergence results in a faster stabilization of system \eqref{eqn:system-case-study}, as visualized in Fig. \ref{fig:convergence-states}.
To accelerate the convergence time in the EC approach, $\alpha$ as a lower bound for $A$ can be increased.
Nevertheless, the convergence time depends on the initial value of the tracking system and the tracking error does not converge in finite time, no matter how $\alpha$ is chosen.

\begin{figure}[thpb]
	\centering
    \includegraphics[width = 0.42\textwidth]{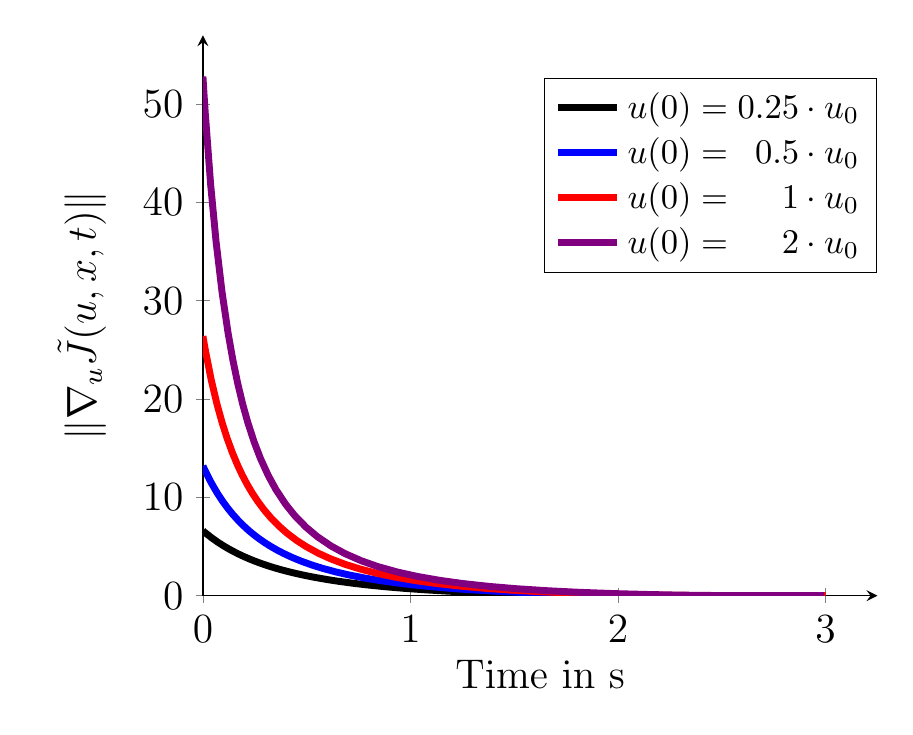}    
    \caption{Convergence of $\nabla_u \tilde J(u,x,t)$ for different initial values.}
    \label{fig:comparison-initial-values}
\end{figure}

To visualize the benefit of tracking system \eqref{eqn:special-choice-h}, the following two modifications are considered:
At first, the convergence of $\norm{\nabla_u \tilde J(u,x,t)}$ is investigated for different initial values in \eqref{eqn:special-choice-h}, namely $u(0) \in \mathbb U_0 = \{ 0.25 \cdot u_0, 0.5 \cdot u_0, 1 \cdot u_0, 2 \cdot u_0 \}$, where $u_0 = \kappa_1(x_0)$ and $x_0 = \begin{pmatrix} -9, & -7, & -5 \end{pmatrix}^\top$.
Since $\varphi(u,x_0) \leq \gamma$ holds for each $u \in \mathbb U_0$, all initial values are feasible.
The results are visualized in Fig. \ref{fig:comparison-initial-values}.
As expected, the choice of $u(0)$ has no influence on the convergence time of $\norm{\nabla_u \tilde J(u,x,t)}$ \ie each tracking error is zero at the specified settling-time $\tau = 3 \text s$.

\begin{figure}[thpb]
	\centering
    \includegraphics[width = 0.42\textwidth]{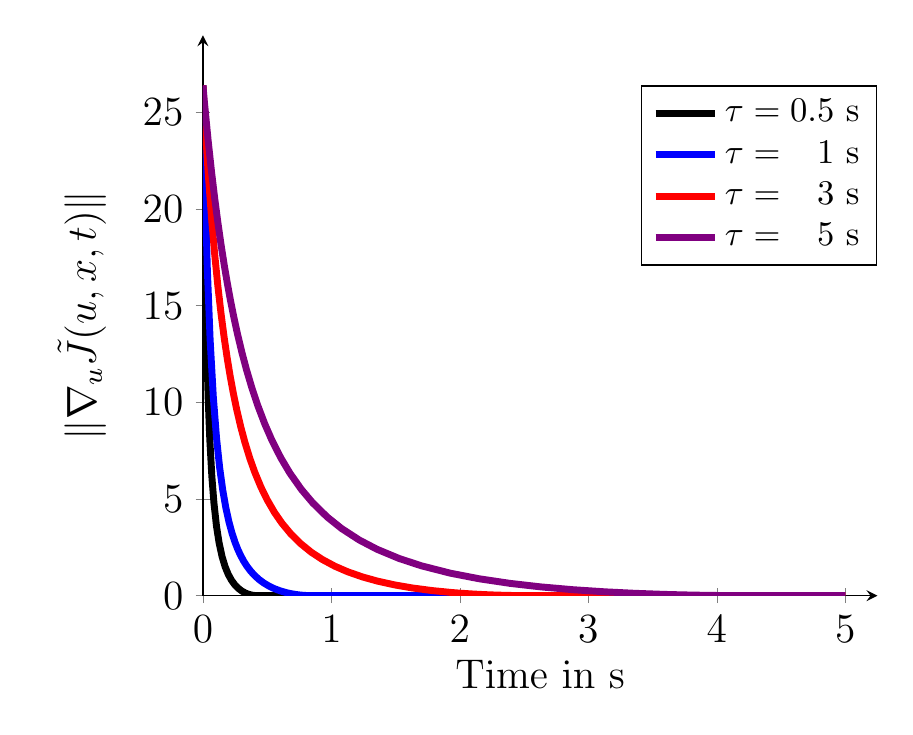}    
    \caption{Convergence of $\nabla_u \tilde J(u,x,t)$ for different settling-times.}
    \label{fig:comparison-settling-time}
\end{figure}

Secondly, the convergence of $\norm{\nabla_u \tilde J(u,x,t)}$ is considered for different settling-times, namely $\tau \in \{ 0.5, 1, 3, 5 \} \text s$.
As it is visible in Fig. \ref{fig:comparison-settling-time}, the gradient of the relaxed objective function $\tilde J$ is zero in each case for $t \geq \tau$, and thus also the tracking error.
However, small settling-times (e.g. $\tau < 0.1 \text s$) result in numerical problems, which are described in Remark \ref{rem:user-defined-settling-time}.
They can be solved up to a certain choice of the settling-time with a smaller step-width of the respective ODE solver.

%%%%%%%%%%%%%%%%%%%%%%%%%%%%%%%%%%%%%%%%%%%%%%%%%%%%%%%%%%%%%%%%%%%%%%%%%%%%%%%%
\section{CONCLUSION AND OUTLOOK} \label{sec:conclusion}

In this paper, a constrained optimization problem with a time-varying cost function that satisfies mild assumptions was considered. 
The Lyapunov function decay constraint was incorporated into the cost function via a barrier function approach, which yields a strongly convex objective function, and thus it exists a unique minimizer.
An ODE was proposed, whose solution converges to zero in a fixed settling-time, which can be chosen by the user.
This ODE was the basis for the design of a tracking system, whose solution converges to the minimizer of the relaxed barrier function in fixed-time.
Furthermore, it was shown that this solution is feasible for all times, and thus it asymptotically stabilizes the input-affine system based on Assumption \ref{asm:decay-condition}.
The effectiveness of the approach was presented in a case study, in which it was compared with an existing method that ensures exponential convergence.
Furthermore, the influence of settling-time and the initial value of the tracking system were investigated.

In future works, the approach can be extended to include input and time-varying constraints.
These two extensions increase the applicability of the proposed approach, since a broader class of problems can be solved.
Lastly, the extension to non-differentiable Lyapunov functions allows to address systems that fail to satisfy Brockett's condition \cite{Brockett1983-stabilization}.

%%%%%%%%%%%%%%%%%%%%%%%%%%%%%%%%%%%%%%%%%%%%%%%%%%%%%%%%%%%%%%%%%%%%%%%%%%%%%%%%%
%\section{ACKNOWLEDGMENTS}
%
%Wir danken ...

\section{APPENDIX}

The following lemma proves that $\tilde J(u,x,t)$ as defined in \eqref{eqn:dfn-of-J-tilde} is strongly convex, and thus the solution of \eqref{eqn:opt-prob} is unique for all $t \in \R_{\geq 0}$.

\begin{lem}[Strongly convex objective function] \label{lem:J-tilde-strongly-convex}
	Consider the objective function $\tilde J(u,x,t)$ of the relaxed problem \eqref{eqn:dfn-of-J-tilde} and let Assumption \ref{asm:J-uni-str-convex} hold. 
	Then, $\tilde J(u,x,t)$ is strongly convex with parameter $m_J$.
\end{lem}

\begin{proof}
	Due to Assumption \ref{asm:J-uni-str-convex}, $J(u,x)$ is strongly convex in $u$ with $m_J$.
	Per definition of a strongly convex function, for all $u_1, u_2 \in \R^m$ it holds that \cite{Boyd2004-convex}
	\begin{equation} \label{eqn:J-strongly-convex}
		J(u_2,x) \geq J(u_1,x) + \nabla_u J(u_1,x)^\top (u_2-u_1) + \frac{1}{m_J} \norm{u_2-u_1}.
	\end{equation}
	Furthermore, a barrier function as per Definition \ref{dfn:barrier-fct} is convex.
	Since $\varphi(u,x) - \gamma$ is affine in $u$, the barrier function has an affine function as argument.
	Therefore, $B(\varphi(u,x) - \gamma)$ as well as $\mu(t) B(\varphi(u,x) - \gamma)$ are convex in $u$ for all $x \in \R^n$ \ie per definition, for all $u_1, u_2 \in \R^m$ it holds that
	\begin{equation} \label{eqn:barrier-convex}
		\begin{split}
			&\mu(t) B(\varphi(u_2,x) - \gamma) \\
			&\geq \mu(t) B(\varphi(u_1,x) - \gamma) + \mu(t) \nabla_u B(\varphi(u_1,x) - \gamma)^\top (u_2-u_1).
		\end{split}
	\end{equation}	
	The sum of \eqref{eqn:J-strongly-convex} and \eqref{eqn:barrier-convex} yields
	\begin{equation} \label{eqn:J-tilde-strongly-convex-proof}
		\begin{split}
			&\tilde J(u_2,x,t) \\
			&= J(u_2,x) + \mu(t) B(\varphi(u_2,x) - \gamma) \\
			&\geq J(u_1,x) + \mu(t) B(\varphi(u_1,x) - \gamma)  + \frac{1}{m_J} \norm{u_2-u_1} \\
			& \ \ \ + (\nabla_u J(u_1,x) + \mu(t) B(\varphi(u_1,x) - \gamma))^\top (u_2-u_1) \\
			&= \tilde J(u_1,x,t) + \nabla_u \tilde J(u_1,x,t)^\top (u_2-u_1) + \frac{1}{m_J} \norm{u_2-u_1}.
		\end{split}
	\end{equation}	
	Since \eqref{eqn:J-tilde-strongly-convex-proof} is the definition of a strongly convex function (cf. \eqref{eqn:J-strongly-convex}), it follows that $\tilde J(u,x,t)$ is strongly convex in $u$ with parameter $m_J$.
\end{proof}

%%%%%%%%%%%%%%%%%%%%%%%%%%%%%%%%%%%%%%%%%%%%%%%%%%%%%%%%%%%%%%%%%%%%%%%%%%%%%%%%

\bibliographystyle{plain}        % Include this if you use bibtex 
\bibliography{bib}

\end{document}